\documentclass[12pt,fleqn]{article}

\usepackage{amsmath,amssymb,graphicx}

\usepackage[textwidth=15.9cm,textheight=23.2cm]{geometry}
\usepackage{latexsym}
\usepackage{cite}
\usepackage{bbm}
\usepackage{mathrsfs}
\numberwithin{equation}{section}

\def\AdSs5{$AdS_5$}
\def\AdSS5{$AdS_5$}
\def\AdS5s5{$AdS_5 \times S^5$}
\def\al{{\alpha^{\prime}}}
\def\gs{g_{st}}

\def\er{{\rm e}}
\def\dr{{\rm d}}

\def\gs{g_{\rm s}}

\newcommand{\be}{\begin{equation}}
\newcommand{\ee}{\end{equation}}
\newcommand{\ba}{\begin{eqnarray}}
\newcommand{\ea}{\end{eqnarray}}
\newcommand{\bdm}{\begin{displaymath}}
\newcommand{\edm}{\end{displaymath}}

\newcommand\fr[1]{\frac{1}{#1}}
\newbox\SlashedBox
\def\fs#1{\setbox\SlashedBox=\hbox{#1}
\hbox to 0pt{\hbox to 1\wd\SlashedBox{\hfil/\hfil}\hss}{#1}}
\def\hboxtosizeof#1#2{\setbox\SlashedBox=\hbox{#1}
\hbox to 1\wd\SlashedBox{#2}}

\def\ms#1{\setbox\SlashedBox=\hbox{$#1$}
\hbox to 0pt{\hbox to 1\wd\SlashedBox{\hfil/\hfil}\hss}#1}

\def\t2{\tau_2}
\def\IZ{\relax\ifmmode\mathchoice {\hbox{\cmss Z\kern-.4em Z}}
{\hbox{\cmss Z\kern-.4em Z}} {\lower.9pt\hbox{\cmsss Z\kern-.4em
Z}} {\lower1.2pt\hbox{\cmsss Z\kern-.4em Z}} \else{\cmss
Z\kern-.4em Z}\fi}

\def\c1{{\chi^1}}

\def\N4{{\cal N}=4}

\def\nn{\nonumber}

\newcommand{\ups}{\upsilon}
\DeclareMathAlphabet{\mathpzc}{OT1}{pzc}{m}{it}
\newcommand\forp[3]{{\it Fortschr.\ Phys.\ }{\bf #1} (#2) #3}

\newcommand\jgp[3]{{\it J.\ Geom.\ Phys.\ }{\bf #1} (#2) #3}
\newcommand\jhep[3]{{\it J. High Energy Phys.\ }{\bf #1} (#2) #3}

\newcommand\npb[3]{{\it Nucl.\ Phys.\ }{\bf B#1} (#2) #3}

\newcommand\prep[3]{{\it Phys.\ Rept.\ }{\bf #1} (#2) #3}

\newcommand{\hepth}[1]{{\tt hep-th/#1}}

\begin{document}

%%%%%%%%%%%%%%%%   TITLE    %%%%%%%%%%%%%%%%%%%%

\thispagestyle{empty}
\renewcommand{\thefootnote}{\fnsymbol{footnote}}

\bigskip\bigskip

\begin{center} \noindent \Large \bf
AdS/CFT Duality and the Higgs Branch \\of ${\cal N}=2$ SYM
\end{center}

\bigskip\bigskip\bigskip

\bigskip\bigskip\bigskip

\centerline{ \normalsize \bf Z. Guralnik$^{\,a}$, S.
Kovacs$^{\,b}$ and B. Kulik$^{\,b}$~\footnote[1]{\noindent \tt
zack@physik.hu-berlin.de, bogdan.kulik@aei.mpg.de,
stefano.kovacs@aei.mpg.de} }

\bigskip
\bigskip\bigskip

\centerline{$^a$ \it Institut f\"ur Physik} \centerline{\it
Humboldt-Universit\"at zu Berlin} \centerline{\it Newtonstra{\ss}e
15} \centerline{\it 12489 Berlin, Germany}
\bigskip
\centerline{$^b$ \it Max-Planck-Institut f\"ur Gravitationsphysik}
\centerline{\it Albert-Einstein-Institut} \centerline{\it Am
M\"uhlenberg 1, D-14476 Golm, Germany}
\bigskip\bigskip

\renewcommand{\thefootnote}{\arabic{footnote}}

\centerline{\bf \small Abstract}
\medskip

\noindent {\small We construct the AdS description of the Higgs
branch of the finite ${\cal N}=2$ $Sp(N)$ gauge theory with one
antisymmetric hypermultiplet and four fundamental hypermultiplets.
Holography, combined with the non-renormalization of the metric on
the Higgs branch, leads to novel constraints on unknown terms in
the non-abelian Dirac-Born-Infeld action.  These terms include
non-minimal couplings of D-branes to bulk supergravity fields.}
\newpage

\section{Introduction}

Many generalisations of the original AdS/CFT duality have been
proposed and in particular there have been numerous
articles discussing supergravity duals of theories with
matter in the fundamental representation, including examples with
confinement and chiral symmetry breaking. We will discuss
the supergravity description of the Higgs branch of a finite
four-dimensional ${\cal N}=2$ gauge theory with fundamental
representations \footnote{In related work, the AdS description of
the Higgs branch of a $(4,4)$ defect CFT was constructed in
\cite{Constable:2002xt}.}. This article is based on work which is
expounded in more detail in \cite{Guralnik:2004ve}.  Besides
allowing one to study the Higgs phase of strongly coupled large
$N$ gauge theories, this work leads to constraints on unknown
terms coupling D-branes to supergravity, as well as possible
cosmological applications \cite{Guralnik:2004wq}.

The AdS description of the Higgs branch
involves a supergravity background with probe D-branes.  We will
specifically consider a finite ${\cal N}=2$ Sp($N$) gauge theory
which is conformal at the origin of moduli space, and is dual to
string theory in AdS$_5 \times S^5/Z_2$,  with D7-branes wrapping
the $Z_2$ fixed subspace with geometry AdS$_5 \times S^3$.  This
background is the near horizon limit of a D3-D7-O7 system, with
the D7-branes treated as probes.

The ${\cal N}=2$ theory we consider has one hypermultiplet in the
antisymmetric representation and four in the fundamental
representation. For this theory, there is a known exact correspondence
between the Higgs branch and the moduli space of Yang-Mills instantons
\cite{Witten:1995gx,Douglas:1996uz} (see \cite{Dorey:2002ik} for a
review). We will show that the equations of motion obtained from the
D7-brane effective action at leading order in the $\alpha'$
expansion admit solutions which are the usual Yang-Mills
instantons, despite the curved background.  The existence of these
solutions is due to a conspiracy between the Yang-Mills and
Wess-Zumino terms in the D7-brane action.

At higher orders in $\alpha'$,  little is known about the non-Abelian
DBI action in flat space, with the exception of a few low order terms
\cite{Tseytlin:1986ti,Gross:1986iv,Tseytlin:1997cs,Koerber:2002zb}.
Even less is known about non-minimal couplings between the
world-volume gauge fields on D-branes and bulk fields (curvature,
$p$-forms and dilaton) which appear at higher orders in $\alpha'$,
although some terms of the form $R^2\,{\rm tr}F^2$ have been studied
\cite{afrey,Wijnholt:2003pw}. We will show that the $F^4$ corrections
in the D-brane effective action do not modify the leading order
solution, but without knowing all the coupling to bulk fields with
non-zero background value we can not explicitly show that conventional
Yang-Mills instantons remain solutions at higher orders in
$\alpha'$. However the exact correspondence between instantons and the
Higgs branch implies that instantons must be solutions to all orders
in the $\alpha'$ expansion. This leads, reversing the point of view,
to constraints on the unknown couplings. We find for example that all
terms containing bulk fields which are quadratic in the D7-brane field
strength must sum to zero when the bulk fields are set equal to their
background values. This constraint is similar in spirit to constraints
on the flat space DBI action which follow from requiring that stable
holomorphic bundles solve the equations of motion
\cite{DeFosse:2001mk,Koerber:2001ka,Koerber:2002zb}.

The metric on the Higgs branch moduli space can be computed at large
't Hooft coupling by considering slowly varying instantons on the
probe D-branes.  The non-renormalization of the metric on the Higgs
branch \cite{Argyres:1996eh} implies that the leading term in the
strong coupling expansion generated by the AdS/CFT duality must be the
only non-zero term, and should equal the weak coupling tree level
result.  We will show that the leading term in our construction gives
rise to the correct metric on the Higgs branch and that higher order
corrections vanish assuming that certain bulk -- brane couplings sum
to zero.

\section{Holography for an ${\cal N}=2$ gauge theory with fundamental
representations}

We will consider the ${\cal N}=2$ theory which describes the low
energy dynamics of a D3-D7-O7 system. This system consists of $4$
D7-branes coincident with an O7-plane,  such that one has a consistent
tadpole free string background, and $N$ D3-branes within the
D7-O7-plane.  It arises considering the near horizon geometry on a
stack of D3-branes in the vicinity of a fixed point in the type IIB
orientifold $T^2/(-1)^F\Omega I$. At low energies, this system is
described by a four dimensional gauge theory with  Sp($N$) gauge
symmetry and SO(8) flavor symmetry.  There is one hypermultiplet in
the anti-symmetric representation and four in the fundamental
representation. The latter arise from open strings stretched between
the D3- and D7-branes, and have non-zero expectation values on
the Higgs branch.

At the origin of the moduli space, the theory is conformal and is
dual to string theory in AdS$_5 \times S^5/Z_2$ with a D7-brane
wrapping the AdS$_5 \times S^3$ fixed surface
\cite{Aharony:1998xz,Fayyazuddin:1998fb}. The near horizon geometry on
the D3-branes is AdS$_5 \times S^5/Z_2$, with metric
\be
\dr s^2 = \frac{r^2}{L^2} \left(-\dr x_0^2 + \dr x_1^2+\dr
x_2^2+\dr x_3^2 \right) + \frac{L^2}{r^2} \left(\dr r^2 + r^2
\dr\hat\Omega_5^2 \right) \, ,
\label{fullD3}
\ee
where $L$ is the radius of both the AdS$_5$ and the $S^5$ factors and
as usual $L^4=4\pi\gs N\al^2$. In (\ref{fullD3}) we have denoted with
$x_\mu$, $\mu=0,1,2,3$, the coordinates on the AdS$_5$ boundary and
with $r$ the radial coordinate transverse to the D3-branes, $r^2 =
X_4^2 + \cdots +X_9^2$.  In (\ref{fullD3}) $\dr\hat\Omega_5^2$ denotes
the metric on $S^5/Z_2$ given by
\be
\dr\hat\Omega_5^2 = \dr\theta^2 +
\sin^2\theta\,\dr\phi^2 + \cos^2\theta \, \dr\Omega_3^2 \, ,
\label{S5Z2metric}
\ee
where the range of $\phi$ is $[0,\pi]$ instead of $[0,2\pi]$ as for an
ordinary $S^5$.

The D7-branes are at a fixed point of the orientifold,
$X_8=X_9=0$. After taking the near horizon limit they fill AdS$_5$
and wrap the $S^3$ corresponding to $\theta=0$ in
(\ref{S5Z2metric}), which is fixed under $Z_2$. The induced metric
on the D7-branes is \be \dr s^2 = \frac{U^2}{L^2} \, \dr
x_\parallel^2 + \frac{L^2}{U^2}\left(\dr U^2 + U^2 \dr\Omega_3^2
\right) = \ups^2 \, \dr x_\parallel^2 + \fr{\ups^2}\,\dr X_\perp^2
\, , \label{mmetric} \ee where \bdm U^2 = r^2\big|_{X_8=X_9=0} =
X_4^2 + X_5^2 +X_6^2 + X_7^2 \edm
 and
\bdm \dr x_\parallel^2 = -\dr x_0^2 + \dr x_1^2 + \dr x_2^2 +\dr
x_3^2 \, , \qquad \dr X_\perp^2 = \dr X_4^2 + \dr X_5^2 +\dr X_6^2
+ \dr X_7^2 \, . \edm For convenience of notation in
(\ref{mmetric}) we have also defined the dimensionless variable
$\ups$ related to $U$ by $\ups^2=U^2/L^2$.

\section{The Higgs branch}

There is a well know exact map between the moduli space of Yang-Mills
instantons and the Higgs branch of the $p+1$ dimensional theories
describing D$p$ -- D($p$+4) brane systems
\cite{Witten:1995gx,Douglas:1996uz}. The Higgs branch corresponds to
D$p$-branes which are not pointlike, but which have instead been
dissolved in the D($p$+4)-branes. Dissolved D$p$-branes can be viewed
as instantons in the $p+5$ dimensional world-volume theory on the
D($p$+4)-branes \cite{Douglas:1995bn}, due to the Wess-Zumino coupling
\begin{equation}
S_{WZ} = \mu_p \int \dr^{p+5}\xi \, C^{(p+1)} \wedge
{\rm tr} \left(F\wedge F\right) \, .
\end{equation}
The low energy degrees of freedom on the D7-branes are described
by an eight-dimensional gauge theory, but due to the curved
geometry resulting from the embedding in the near horizon geometry
(\ref{fullD3}) of the D3-branes the existence of instanton
solutions in such a theory is far from obvious. Moreover the
inclusion of higher order corrections gives rise to an infinite
number of higher dimensional couplings which could modify the
solutions of the leading order equations of motion. We will find
however that despite the curved geometry the theory admits
ordinary instanton solutions which under certain assumptions are
not corrected by the inclusion of higher derivative interactions.
The D7 action takes the form,
\begin{align}
\label{dbig} S = &\frac{1}{(2\pi)^7 \gs\alpha'^4} \int\, \sum_q
C^{(q)} \wedge\, {\rm tr}\, ( \er^{2\pi\alpha' F}) \nonumber
\\ &+ \frac{1}{(2\pi)^7 \gs\alpha'^4} \int \dr^8 x\, \sqrt{-g}
\,(2\pi\alpha')^2 \,\frac{1}{4}\,{\rm tr} (F_{AB}F^{AB})+\cdots \,
.
\end{align}
We have not  written terms involving world-volume fermions or
scalars. This action is the sum of a Wess-Zumino term, a
Yang-Mills term, and an infinite number of $\alpha'$ corrections
represented by ``$\cdots$''. Very little is known about the
latter. Nevertheless, the correspondence between instantons and
the Higgs branch suggest that the equations of motion should be
solved by field strengths which are self dual with respect to a
flat four-dimensional metric.

Let us first consider the equations of motion to leading order in
the $\alpha'$ expansion\footnote{In the AdS setting, the $\alpha'
$ expansion effectively becomes a large 't Hooft coupling
expansion.}. With non-trivial field strengths only in the
directions $X_\perp$, the leading order action for D7-branes
embedded in (\ref{fullD3}) with induced metric (\ref{mmetric}) is
\begin{align}\label{nerg}
S &= \frac{1}{(2\pi)^5 \gs\alpha'^2} \left( \int\, C^{(4)}
\wedge\, {\rm tr}\, F\wedge F +  \int \dr^8 x\, \sqrt{-g} \,
\frac{1}{4}\,{\rm tr} (F_{ab}F^{ab})\right)  \nonumber \\
&= \frac{N}{(2\pi)^4\lambda L^4} \int \dr^4 x_\parallel \, \dr^4
X_\perp \, \ups^4 \, \frac{1}{2}\,{\rm
tr}(\frac{1}{2}\epsilon_{mnrs}F_{mn}F_{rs} + F_{mn}F_{mn}) \nn \\
&=\frac{N}{(2\pi)^4\lambda L^4} \int \dr^4 x_\parallel \, \dr^4
X_\perp \, \ups^4 \, \frac{1}{4}\,{\rm tr}F_{+}^2 \, ,
\end{align}
where the lowercase latin indices $m$ label the  $X_\perp =
X^{4,5,6,7}$ directions and, to arrive at the last line, we have
used the explicit form of the Ramond-Ramond four-form in AdS$_5
\times S^5/Z_2$,
\begin{align}
C^{(4)}_{0123} = \frac{U^4}{L^4} \, .
\end{align}
Thus, at leading order in $\alpha'$, field strengths for which
$F^+ = 0$ (anti-self-dual with respect to the flat metric
$\dr X_\perp^2$) solve the equations of motion due to a conspiracy
between the Wess-Zumino and Yang-Mills terms.

The correspondence with the Higgs branch moduli space of the
${\cal N}=2$ SYM theory requires that the $F^+=0$ configurations
remain solutions when higher order corrections are included in the
D7-brane effective action.  Temporarily neglecting terms which
involve non-minimal couplings to bulk supergravity fields, the
action to order $\alpha'^2$ is given by
\begin{equation}\label{fonly}
S = \frac{N}{\lambda^2}\frac{1}{(2\pi)^5}\int\, {\rm tr}\,
\left[\frac{1}{2} x_\perp^4 F^+_{mn}F^+_{mn}
 -\frac{x_\perp^8}{4\pi\lambda}\frac{1}{384}
\left(2 F^+_{mn}F^+_{mn}F^-_{rs}F^-_{rs} +
F^+_{mn}F^-_{rs}F^+_{mn}F^-_{rs} \right)\right]\, ,
\end{equation}
where we have written the (known
\cite{Gross:1986iv,Tseytlin:1986ti,Tseytlin:1997cs}) $F^4$ terms
in terms of self dual and anti-self-dual field strengths.  Since
the $F^4$ terms are quadratic in $F^+$, anti-self-dual field
strengths are still manifestly solutions of the equations of
motion of (\ref{fonly}). However, even to this order in the
$\alpha'$ expansion, not all the couplings to bulk fields needed
for a complete proof are known. There may also be terms of the
general form $R^2 {\rm tr} F^2, {\cal F}_{(5)}^4 {\rm tr}F^2, R
{\cal F}_{(5)}^2{\rm tr}F^2$  which effect the equations of motion
in the AdS background,  for which the curvature $R$ and
Ramond-Ramond five-form ${\cal F}_{(5)}$ are non-vanishing. Rather
than proving that self-dual field strengths solve the equations of
motion, we will show that the existence of such solutions leads to
constraints on the unknown couplings.

The CP odd Wess-Zumino term proportional to $\upsilon^4
\epsilon_{mnrs}F_{mn}F_{rs}$  is exact, with no corrections at any
order in $\alpha'$. In order to preserve the $F^+ = 0$ solutions,
the quadratic CP even term must be $\upsilon^4\, \frac{1}{2}\,
{\rm tr} F_{mn}F_{mn}$ with exactly the same coefficient. As
discussed above, this is already the case at leading order in
$\alpha'$. Thus, at every order in the $\alpha'$ expansion,  the
terms of the form $f(R,{\cal F}_{(5)}){\rm tr} F^2$ must sum to
zero when the bulk fields are set equal to their AdS values.

Some terms of the form $R^2 F^2$ have appeared in the literature.
These are \cite{afrey,Wijnholt:2003pw}
\begin{align}\label{cterms}S_{R^2F^2} &= -\mu_p (2\pi\alpha')^2 \int
\sqrt{g}\, \frac{1}{4}{\rm tr}F_{\alpha\beta}F^{\alpha\beta}
\left[\frac{1}{24}\frac{(4\pi^2\alpha')^2}{32\pi^2}
\left((R_T)_{\alpha\beta\gamma\delta}(R_T)^{\alpha\beta\gamma\delta}
\right.\right.\nonumber\\
&\left.\left. \qquad\qquad\qquad- 2 (R_T)_{\alpha\beta}
(R_T)^{\alpha\beta} -(R_N)_{ab\alpha\beta}(R_N)^{ab\alpha\beta}+
2\bar R_{ab}\bar R^{ab}\right) \right] \, .
 \end{align}
The curvature terms appearing after
$F_{\alpha\beta}F^{\alpha\beta}$ are the same as the pure $R^2$
terms computed in \cite{Bachas:1999um}.  The various curvature
tensors appearing in (\ref{cterms}) are defined in
\cite{Bachas:1999um}. For the special case of an embedding with
vanishing second fundamental form, the tensors
$(R_T)_{\alpha\beta\gamma\delta}$ and $(R_N)_{ab\alpha\beta}$ are
just pull-backs of the bulk Riemann tensor to the tangent and
normal bundle, indicated by greek and latin indices respectively
(we emphasize that this is a change of notation from the previous
sections). The tensors $(R_T)_{\alpha\beta}$ and $\bar R_{ab}$ are
not pull-backs of the bulk Ricci tensor, but are obtained from
contractions of tangent indices in the pull-backs of the Riemann
tensor. Specifically, for vanishing second fundamental form,
\begin{align}
\bar R_{ab} \equiv g^{\alpha\beta}R_{\alpha ab \beta},\qquad (
R_T)_{\alpha\beta} \equiv g^{\lambda\mu}
R_{\lambda\alpha\mu\beta}\, ,
\end{align}
where $g_{\alpha\beta}$ is the induced metric on the D-brane. For
the AdS$_5 \times S^3$ embedding in AdS$_5 \times S^5/Z_2$, the second
fundamental form vanishes. In this background,
\begin{equation}
(R_T)_{\alpha\beta\gamma\delta}(R_T)^{\alpha\beta\gamma\delta} -
2(\hat R_T)_{\alpha\beta} (\hat R_T)^{\alpha\beta} -
(R_N)_{ab\alpha\beta}(R_N)^{ab\alpha\beta}+ 2\bar R_{ab}\bar
R^{ab} = -\frac{6}{25} L^2 \, ,
\end{equation}
where $L^2 = \sqrt{\lambda}\al$ is the square of the AdS$_5$ (or
$S^5$) curvature radius. Thus (\ref{cterms}) can not be the only term
of the form $f(R, {\cal F}^{(5)}) F^2$ at order $\alpha'^2$, which
must collectively sum to zero in the AdS background.

\section{The metric on the Higgs Branch}

To two derivative order, the effective action on the Higgs branch
of the four-dimensional ${\cal N}=2$ theory we are considering is
equivalent to the action describing slowly varying ``instantons''
in eight-dimensional super Yang-Mills (see \cite{Dorey:2002ik} for
a review). This action has the form
\begin{equation}
S= \int \dr x^0 \cdots \dr x^3 G_{ij}({\cal M})\partial_\mu
{\cal M}^i \partial^\mu {\cal M}^j \, ,
\label{modspapprox}
\end{equation}
where ${\cal M}_i(x^\mu)$ are either Higgs branch or instanton moduli.
From the point of view of the eight-dimensional theory the instantons
we are considering are solitons for which the gauge fields only depend
on the four Euclidean coordinates  $x^{4,5,6,7}$ and have a self-dual
field strength with respect the flat metric in these directions. These
solutions depend on moduli ${\cal M}_i$. The metric (\ref{modspapprox})
is obtained in the moduli space approximation in which the parameters
${\cal M}_i$ are allowed to depend on the coordinates $x^{0,1,2,3}$,
but are slowly varying. The metric $G_{ij}({\cal M})$ is also known to
be tree level exact in the four-dimensional ${\cal N}=2$ theory.

In the AdS dual description of the ${\cal N}=2$  theory, one can
compute the metric on the Higgs branch by finding the action for
slowly varying instantons of the D7-brane theory (\ref{dbig}).
To two derivative order, the effective action must be the same as
that for slowly varying instantons in conventional super
Yang-Mills theory in eight-dimensional flat space, which gives the
exact un-renormalized metric on the Higgs branch.

The metric on the Higgs branch is determined by inserting the
instanton solution into the action, letting the moduli depend on the
coordinates $x^0,\ldots,x^3$, which we indicate by greek indices. The
instantons are localized in the directions $x^m = x^{4,5,6,7}$ and
depend on moduli ${\cal M}_i$, $A_m = A_m^{\rm inst}(x^n, {\cal
M}_i)$. Configurations in which the moduli are coordinate dependent,
$A_m = A^{\rm inst}_m(x^n, {\cal M}_i(x^\alpha))$, are approximate
solution in the limit of slowly varying moduli. More precisely,  the
metric on the Higgs branch can be extracted from the equations which
configurations
\begin{align}
\label{config}
A_m &= A^{\rm inst}_m(x^n, {\cal M}_i(x^\alpha))
\nonumber \\
A_\mu &= \Omega_i \partial_\mu  {\cal M}_i(x^\alpha)
\end{align}
must satisfy in order to solve the full equations of motion to
leading (two derivative) order in a derivative expansion.  The
relevant terms in the Dirac Born Infeld action are those involving
two greek indices.  Two order $\alpha'^2$ (or equivalently ${\cal
O}(1/\lambda)$), the relevant terms are (neglecting bulk
couplings)
\begin{align}
S & = \frac{N}{\lambda} \frac{1}{(2\pi)^5} \int \dr^4
X_\perp\,\frac{1}{4} {\rm tr}(F_{\mu m} F_{\mu m})  \nonumber\\
&+\frac{N}{\lambda^2}\frac{1}{(2\pi)^6}\int\, \dr^4
X_\perp\,X_\perp^4\,\frac{1}{12}{\rm tr} \left[F_{s\mu}F_{\mu
n}\left(\{F_{nr},F_{rs}\}
-\frac{1}{2}\delta_{sn}F_{tu}F_{ut}\right)\right. \nonumber\\
&+\frac{1}{2}\left. \left(F_{\mu n}F_{nr}F_{s\mu}F_{rs} + F_{\mu
n}F_{rs}F_{s\mu}F_{nr} - \frac{1}{2}F_{\mu
n}F_{rs}F_{n\mu}F_{sr}\right)\right] \,.
\end{align}
Note that the leading term is just that of Yang-Mills theory in
eight flat dimensions; the warp factors appearing in the $AdS^5
\times S^3$ metric cancel in this term.   It is useful to rewrite
the field strengths $F_{mn}$ in the subleading term in terms of
self-dual and anti-self-dual parts, giving
\begin{align}
\label{metact}
S & = \frac{N}{\lambda} \frac{1}{(2\pi)^5} \int \dr^4
X_\perp\,\frac{1}{4} {\rm tr}(F_{\mu m} F_{\mu m})  \nonumber\\
&+\frac{N}{\lambda^2}\frac{1}{(2\pi)^6}\int\, \dr^4
X_\perp\,X_\perp^4\,\frac{1}{48}{\rm tr} \left[F_{s\mu}F_{\mu
n}\left(\{F^+_{nr},F^-_{rs}\}+\{F^-_{nr},F^+_{rs}\}
\right)\right.  \nonumber\\
&+\frac{1}{2}\left. \left(F_{\mu n}F^+_{nr}F_{s\mu}F^-_{rs}+ F_{\mu
n}F^-_{nr}F_{s\mu}F^+_{rs} + F_{\mu n}F^+_{rs}F_{s\mu}F^-_{nr}+
F_{\mu n}F^-_{rs}F_{s\mu}F^+_{nr} \right)\right] \, .
\end{align}
The equations of motion $\frac{\delta}{\delta A_\mu} S =0$ give
\begin{align}
\partial_\mu {\cal M}_i\left(D_m \frac{\delta A_m}{\delta {\cal M}_i}
- D_m D_m \Omega_i\right) = 0\, ,
\end{align}
which has a unique solution for $\Omega_i$ as a function of $x^m$ and
${\cal M}_i$. Taking this $\Omega_i$ and inserting (\ref{config})
into the action (\ref{metact}) gives the metric on the Higgs branch
via the relation
\begin{align}
\partial_\mu{\cal M}^i \partial_\mu {\cal M}^j G_{ij}({\cal M}) =
\frac{N}{\lambda} \frac{1}{(2\pi)^5} \int d^4 X_\perp\,\frac{1}{4}
{\rm tr}(F_{\mu m} F_{\mu m}) \, ,
\end{align}
where the higher order term vanishes because the configuration
(\ref{config}) satisfies $F^+ =0$. One therefore gets the same
result from instantons on the D7-brane embedded in AdS as one gets
from Yang-Mills theory in flat space.  We emphasize that this
result assumes that bulk couplings of the form $h_{nm\mu\nu}({R},
{\cal F}^{(5)})F_{\mu n} F_{\nu m}$ sum to zero when ${R}$ and
${\cal F}^{(5)}$ are set to their AdS background values. With this
assumption,  the non-renormalization of the metric on the Higgs
branch is realized in the strong coupling expansion obtained using
holography.  The leading and only term is the same as the exact
tree level result.

\end{document}